\documentclass[conference,10pt,final]{IEEEtran}

\usepackage[T1]{fontenc}
\usepackage[english]{babel}
\usepackage[utf8]{inputenc}
\usepackage{amsmath}
\usepackage{amsfonts}
\usepackage{amssymb}
\usepackage{balance}
\usepackage{graphicx}
\usepackage{epstopdf}
\usepackage{epsfig}
\usepackage{tikz}
\usepackage{soul}
\usetikzlibrary{shapes,arrows}

\usepackage{algorithm}
\usepackage{color}
\usepackage{pifont}
\usepackage{verbatim,mathrsfs, cite}
\usepackage[acronym,shortcuts]{glossaries}
\usepackage{setspace, url}
\usepackage{bm}

\makeglossaries

\newacronym{AF}{AF}{amplify and forward}
\newacronym{AN}{AN}{artificial noise}
\newacronym{AWGN}{AWGN}{additive white Gaussian noise}
\newacronym{BS}{BS}{base station}
\newacronym{CDF}{CDF}{cumulative distribution function}
\newacronym{CF}{CF}{compute and forward}
\newacronym{CSI}{CSI}{channel-state information}
\newacronym{DF}{DF}{decode and forward}
\newacronym{FD}{FD}{full-duplex}
\newacronym{HD}{HD}{half-duplex}
\newacronym{iid}{i.i.d.}{independent and identically distributed}
\newacronym{MAC}{MAC}{multiple access channel}
\newacronym{MIMO}{MIMO}{multiple-input multiple-output}
\newacronym{MIMOME}{MIMOME}{MIMO multiple-Eve}
\newacronym{MT}{MT}{mobile terminal}
\newacronym{PDF}{PDF}{probability density function}
\newacronym{PLS}{PLS}{physical layer security}
\newacronym{RB}{RB}{random binning}
\newacronym{SCF}{SCF}{scaled compute-and-forward}
\newacronym{SOP}{SOP}{secrecy outage probability}
\newacronym{SVD}{SVD}{singular value decomposition}
\newacronym{SNR}{SNR}{signal to noise ratio}

\title{Secure Compute-and-Forward Transmission With Artificial Noise and Full-Duplex Devices}
\author{
	\IEEEauthorblockN{Stefano Tomasin}
	\IEEEauthorblockA{Department of Information Engineering \\ University of Padova, Italy
}}

\begin{document}
\maketitle

\begin{abstract}
We consider a wiretap channel with an eavesdropper (Eve) and an honest but curious relay (Ray). Ray and the destination (Bob) are \ac{FD} devices. Since we aim at not revealing information on the secret message to the relay, we consider the \ac{SCF} where scaled lattice coding is used in the transmission by both the source (Alice) and Bob in order to allow Ray to decode only a linear combination of the two messages. At the same time Ray transmits \ac{AN} to confuse Eve. When Ray relays the decoded linear combination, Alice and Bob are transmitting \ac{AN} against Eve. This can be a 5G cellular communication scenario where a \ac{MT} aims at transmitting a secret message to a \ac{FD} \ac{BS}, with the assistance of a network \ac{FD} relay. With respect to existing literature the innovations of this paper are:  a) Bob and Ray are \ac{FD} devices; b) Alice, Ray and Bob transmit also \ac{AN}; and c) the channel to Eve is not known to Alice, Bob and Ray. For this scenario we derive bounds on both the secrecy outage probability under Rayleigh fading conditions of the channels to Eve, and the achievable secrecy-outage rates. 
\end{abstract}


\glsresetall

\section{Introduction}

The next generation of mobile communication systems (5G) will most probably encompass various technological innovations, including among others, \ac{FD} devices \cite{Zhang-2015} and multi-hop (relayed) transmissions. It has also been advocated \cite{Yang} that security should be extended at the physical layer using, as a complement to traditional computational security, \ac{PLS} approaches. 

In this paper we focus on \ac{PLS} solutions for a relay-assisted communication, where both the relay and the destination are \ac{FD} devices. It has already been shown the advantage of cooperation (relaying) for physical layer security \cite{Wang-2015}, which motivates the focus on this approach. This scenario arises for example in a cellular 5G system, where a \ac{MT} aims at transmitting a secret message to a \ac{FD} \ac{BS}, with the assistance of a network \ac{FD} relay. This scenario has been considered in the literature: for example, for \ac{FD} relay see \cite{XChen-2015, Rodriguez-2015} and references therein. In \cite{HeYener-08} an un-authenticated relay was considered (to be kept in the darkness of the secret message), and the destination sent \ac{AN} during the source's transmission, and both an upper bound on the secret rate and achievable rates were derived. When the relay is secure, linear precoding schemes can be applied on a \ac{DF} transmission with \ac{AN} \cite{Huang11}. \ac{AF} relaying and simultaneous jamming by other nodes has been considered in \cite{Chen-11} where selection of relay and jamming nodes has been addressed. 

The case of \ac{FD} destination without relaying is considered in \cite{Li-2012} in the absence of \ac{CSI} on the eavesdropper channel, and in  \cite{Zheng-13} under various \ac{CSI} assumptions on the legitimate and eavesdropper channels. The case of a communication between single-antenna devices assisted by a multi-antenna \ac{FD} relay is considered in \cite{Zhu-2014}, where  joint information and jamming beamforming are designed to guarantee secrecy. A \ac{DF} solution for \ac{FD} relaying is also considered in \cite{Chen-2015} showing the advantages of \ac{FD} over \ac{HD} network solutions. The multi-source multi-relay scenario has been considered in various papers (see \cite{Fan-2016} and references therein), with and without \ac{AF}/\ac{DF} and various knowledge of the \ac{CSI} of the eavesdropper channel, and selection of sources and relays have been optimized. In \cite{Poor-2016} a modulo-and-forward is considered for a single relay without eavesdropper, and in the case of no \ac{CSI} the outage probability is derived. In \cite{Ren-2016} a \ac{SCF} was introduced, and the presence of an eavesdropper was also considered. However the source and destination nodes perfectly know the \ac{CSI} to the eavesdropper, which is not always a realistic assumption.

In this paper we consider that both Ray and Bob are \ac{FD}. Since we aim at not revealing information on the secret message to the relay, we consider the \ac{SCF}. At the same time Ray transmits \ac{AN} to confuse Eve. When Ray relays the decoded linear combination, Alice and Bob are transmitting \ac{AN} against Eve. With respect to existing literature the innovations of this paper are:  a) Bob and Ray are \ac{FD} devices; b) Alice, Ray and Bob transmit also \ac{AN}; and c) the channel to Eve is not known to Alice, Bob and Ray. For this scenario we derive a bound on the secrecy outage probability. In particular, for the case of Rayleigh fading conditions of the channels to Eve, we derive a close form expression of the secrecy outage probability bound.

\section{System Model}

We consider a relay network with a device (Alice) willing to convey a message to another device (Bob), with the help of a relay (Ray). The message must remain secret to a potential passive eavesdropper (Eve), whose location (and even existence) is unknown to both Alice and Bob. Ray is always honest, thus he complies with the transmission protocol rules and aims at supporting Alice and Bob at best. However, he is curious, i.e., willing to know the secret message: thus the communication protocol ensures that the message remains secret also to him. In any case, Eve and Ray are not colluding to get information on the secret message.

Both Alice and Eve\footnote{Since we are considering Eve as a {\em passive} eavesdropper, nothing changes if we assume that she is \ac{FD}.} are assumed to be \ac{HD} devices, while both Bob and Ray have \ac{FD} capabilities. All users are equipped with a single antenna. The devices have a maximum unitary transmit power, and we assume that each receiver is subject to a zero-mean unitary-power \ac{AWGN}. Extension to the case of devices equipped with multiple antennas is left for future study.

We assume that there is no direct link between Alice and Bob, therefore the support of Ray is needed. Among antennas we have flat, reciprocal and \ac{AWGN} channels. So that the complex channel between users are for the Alice/Bob-Ray channel $G_{\rm x}$ with x = A or B, and for Alice/Bob/Ray-Eve channel $Q_{\rm x}$ with x = A or B or R. Since we assume that the relay is curious but honest, we assume that he let Alice and Bob perfectly know the correct channel gains. On the other hand, Eve is not honest and we only assume to know the statistics of its channel to both Alice and Bob.

\section{Secure Communication Protocols}

The communication protocol  is based on \ac{SCF}. Starting from a lattice $\Lambda$, we construct two lattices $\Lambda_{\rm A}, \Lambda_{\rm B} \subset \Lambda$, having unitary second moment, being good in both quantizing and shaping sense of \cite{Erez-2004}. Let $\mathcal V_{\rm A}$ and $\mathcal V_{\rm B}$ be the fundamental Voronoi region of $\Lambda_{\rm A}$ and $\Lambda_{\rm B}$, respectively. Let also $a_1, a_2 \in \mathbb{Z}$, and $\beta_{\rm A}, \beta_{\rm B} \in \mathbb{R}^+$.

\begin{figure}
\centering
\begin{tabular}{c@{\hskip .5in}c}
\begin{tikzpicture}[every path/.style={>=latex},every node/.style={draw,circle}]
  \node            (a) at (0,-1)  { A };
  \node            (b) at (1,0)  { R };
  \node            (c) at (2,-1) { B };
  \node            (d) at (1,-2) { E };  
  \draw[->] (a) edge (b);
  \draw[->] (a) edge (d);
  \draw[->] (c) edge (b);
  \draw[->] (c) edge (d);  
  \draw[->, dashed] (b) edge (d);    
\end{tikzpicture} &
\begin{tikzpicture}[every path/.style={>=latex},every node/.style={draw,circle}]
  \node            (a) at (0,-1)  { A };
  \node            (b) at (1,0)  { R };
  \node            (c) at (2,-1) { B };
  \node            (d) at (1,-2) { E };  
  \draw[->,dashed] (a) edge (d);
  \draw[->] (b) edge (c);
  \draw[->] (b) edge (d);  
  \draw[->, dashed] (c) edge (d);    
\end{tikzpicture} \\
a) Phase 1 & b) Phase 2
\end{tabular}
\caption{Transmissions in the two phases of the proposed scheme among Alice (A), Bob (B), Ray (R) and Eve (E). Solid lines represent transmission of messages, while dashed lines represent \ac{AN} transmissions.}
\label{phases}
\end{figure}
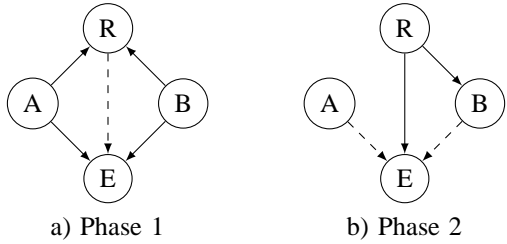

The protocol comprises two phases, whose transmission activities are reported in Fig. \ref{phases}. Solid lines represent transmission of messages, while dashed lines represent transmissions of \ac{AN}. In both phases Eve  listens to ongoing transmissions. The other devices operate as follows.\\

\noindent {\em First phase:}
\begin{itemize}
\item Alice encodes message $\mathcal M_{\rm A}$ of rate $R_{\rm S}$ by applying the random binning approach of \cite{Ren-2016} and selecting a lattice vector $S^A \in \Lambda \cap \mathcal V^A$. Then she transmits the lattice vector $X^A = [S^A/\beta_{\rm A} + D^A] \pmod{\Lambda_{\rm A}/\beta_{\rm A}}$, where $D^A$ is dither uniformly chosen from the scaled Voronoi region $\mathcal V_{\rm A}/\beta_{\rm A}$. Let $R_0$ be the random signal rate and 
\begin{equation}
R_{\rm A} = R_0+R_{\rm S}
\label{Ra}
\end{equation}
the rate of $S^A$.
\item Bob encodes a random message with rate $R_{\rm B}$ $V^B \in \Lambda \cap \mathcal V^B$  chosen uniformly at random, and the transmitted vector is $X^B = [V^B/\beta_{\rm B} + D^B] \pmod{\Lambda_{\rm B}/\beta_{\rm B}}$, where $D^B$ is dither uniformly chosen from the scaled Voronoi region $\mathcal V_{\rm B}/\beta_{\rm B}$. 
\item Ray is receiving the signal from Alice and Bob and at the same time is transmitting zero-mean Gaussian \ac{AN}.
\end{itemize}

\noindent {\em Second phase:}
\begin{itemize}
\item Alice transmits \ac{AN}.
\item Ray decodes 
\begin{equation}
U = a_1 S^A + a_2V^B
\label{Ur}
\end{equation}
(conditions for successful decoding will be discussed in the following section), and computes $\tilde{U} = U/a_1 \pmod{\Lambda^A}$. Then he encodes $\tilde{U}$ with a secrecy capacity achieving code using a random message with rate $R_{\rm R}$, and transmits the encoded message $X^R$ with rate $R_{\rm R} + R_{\rm A}$.
\item Bob receives the message from Ray, and at the same time transmits zero-mean Gaussian \ac{AN}.
\end{itemize}

We want Bob to decode the secret message at the end of the two phases, by letting at the same time Ray and Eve in the darkness of the secret message. The purpose of transmissions of message $\mathcal M_{\rm B}$ and \acp{AN} is to obtain secrecy.  Ray, beyond operating according to the protocol, will also try to get information on the secret message. Note that the \ac{AN} generated by Alice in the second phase does not affect Bob, since Bob does not receive signals from Alice. 

This protocol is a generalization of the protocol described in \cite{Ren-2016} for the following reasons: a) Bob and Ray are \ac{FD} devices; b) Alice, Ray and Bob transmit also \ac{AN}; and c) the channel to Eve is not known to Alice, Bob and Ray.

In the following we assume that the achievable rate of the resulting Ray-Bob channel is higher than that of the Alice-Ray channel, in the absence of Bob transmissions\footnote{Generalization to any ratio of power is possible, with more elaborate expressions.}.

\subsection{Decodability Conditions}

We first consider here the conditions on the various parameters ($a_1, a_2, \beta_{\rm A}$ and $\beta_{\rm B}$) and rates ($R_{\rm A}$ and $R_{\rm B}$) that ensure decodability of the secret message $S^A$ at Bob. Let 
\begin{equation}
\delta_{\rm A} = |G_{\rm A}|^2 \,, \quad \delta_{\rm B} = |G_{\rm B}|^2\,,
\end{equation}
be the \acp{SNR} of the Alice-Ray and Bob-Ray channel, respectively. 

For decodability with vanishing error probability at Ray of $U$ with infinite codeword length we must have \cite{Ren-2016} 
\begin{equation}
R_{\rm A} \leq \log_2 \left(\frac{\beta_{\rm A}^2 \delta_{\rm A}}{M_{\rm N}}\right)\,, \quad R_{\rm B} \leq \log_2\left(\frac{ \beta_{\rm B}^2 \delta_{\rm B}}{M_{\rm N}}\right)
\label{decod}
\end{equation}
where
\begin{equation}
\begin{split}
M_{\rm N} = & \frac{\delta_{\rm A}\delta_{\rm B} (a_1 \beta_{\rm A} - a_2 \beta_{\rm B})^2 + (a_1 \beta_{\rm A})^2\delta_{\rm A}^2 + (a_2\beta_{\rm B})^2 \delta_{\rm B}}{\delta_{\rm A} + \delta_{\rm B} +1}\,.
\end{split}
\end{equation}
For decodability of the secret message sent by Ray in the second phase we must ensure
\begin{equation}
 \log_2  (1 +  \delta_{\rm B}^2) \geq R_{\rm R} + R_{\rm A}\,.
\end{equation}

\section{Achievable Secrecy Rate}

We now consider the secrecy conditions for the proposed protocol. We will first consider the case without Eve, where we only want to let Ray not to get any information on $\mathcal M_{\rm A}$, and then consider the case in which Eve is also present and we want to let her too in the darkness of the secret message.

\subsection{Achievable Secrecy Rate Without Eve}

We assume now that Eve is not present. Then, the protocol must ensures perfect secrecy only with respect to Ray.

The resulting system corresponds to the scenario of \cite{Ren-2016} for which the achievable secrecy rate is bounded as 
\begin{equation}
R_S \leq \log_2 \frac{1 + \delta_{\rm A} + \delta_{\rm B}}{[\sqrt{(1 + \delta_{\rm A})(1+\delta_{\rm B})} - \sqrt{\delta_{\rm A}\delta_{\rm B}}]^2} - 2\,,
\label{Rs}
\end{equation}
with the maximum achieved with $a_1 = a_2 = 1$ and 
\begin{equation}
\frac{\beta_1}{\beta_2} = \sqrt{\frac{\delta_{\rm B} (1 + \delta_{\rm A})}{\delta_{\rm A} (1+\delta_{\rm B})}}\,.
\end{equation}

\subsection{Achievable Secrecy Rate With Eve}

Since here we assume that legitimate users only have a statistical description of the channel to Eve, we cannot employ the approach of \cite{Ren-2016} that imposed that a linear combination of messages $\mathcal M_{\rm A}$ and $\mathcal M_{\rm B}$ was also decodable by Eve. Moreover, we should also take into account the fact that Eve receives the message sent by Ray.

About Eve, we only aim at preventing her from getting information on the secret message, not caring if she decode some liner combination of $\mathcal M_A$ and $\mathcal M_B$. Therefore two actions will be taken:
\begin{itemize}
\item in the first phase Ray transmits \ac{AN} that will lower the decoding capabilities of Eve, and 
\item in the second phase Ray encodes with a random binning approach his message.
\end{itemize}
These two actions will not be enough to guarantee that Eve does not get any information on the secret message. There will be in any case an outage event, i.e., the event in which Eve gets some information on $\mathcal M_{\rm A}$. We can upper-bound this outage probability by computing the probability that either Eve is able to get some information from either the first phase or the second phase. Let $\mathcal O_i$ be the outage event in phase $i=1,2$. We will derive a superset of the outage event in the first phase, i.e., $\mathcal O_1 \subseteq \bar{\mathcal O}_1$, and we will upper-bound the outage probability as
\begin{equation}
P_{\rm out} = 1 - {\rm P}[\mathcal O_1^C, \mathcal O_2^C] \leq 1 - {\rm P}[(\bar{\mathcal O}_1)^C, \mathcal O_2^C]\,, 
\label{Pout}
\end{equation}
where $\mathcal A^C$ represents the complementary event to event $\mathcal A$.

\paragraph*{Remark} We will ensure that in the second phase Eve does not get information on both $\mathcal M_{\rm A}$ and $\mathcal M_{\rm B}$, since a leakage on $\mathcal M_{\rm B}$ could help her in extracting information on $\mathcal M_{\rm A}$ from what she received in the first phase. 

\paragraph*{Remark} We still need to ensure that Ray does not get any information on the secret message, hence (\ref{Rs}) must still hold.

We now detail the actions and the corresponding secrecy outage probabilities. 

\paragraph*{First Phase} Let $Y^E$ be the signal received by Eve in the first phase. The information leakage rate on $\mathcal M_{\rm A}$ to Eve is 
\begin{equation}
\begin{split}
R_{L} = & \frac{1}{N} I(\mathcal M_{\rm A}; Y^E) \\
= & \frac{1}{N} \left[ H(\mathcal M_{\rm A}) - H(X^A, X^B| Y^E)\right. \\
 - H(&  \left.\mathcal M_{\rm A}|Y^E, X^A, X^B) + H(X^A, X^B|Y^E, \mathcal M_{\rm A}) \right] \\
= & \frac{1}{N} \left[ H(\mathcal M_{\rm A}) - H(X^A, X^B) + I(X^A, X^B; Y^E) \right.\\
+   H(&\left.X^B) - I(X^B; Y^{E'})\right] 
\end{split}
\end{equation}
where $H(\cdot)$ and $I(\cdot;\cdot)$ are the entropy and the mutual information, and in the second line we observe that if Eve knows $\mathcal M_{\rm A}$ she can subtract $X^A$ from $Y^E$ (to obtain $Y^{E'}$), and $H(X^A, X^B|Y^E, \mathcal M_{\rm A}) = H(X^B|Y^E)$ (since by knowing $\mathcal M_{\rm A}$ she also knows $X^A$). We thus obtain the upper bound  
\begin{equation}
R_{L}  < \frac{1}{N} \left[  I(X^A, X^B; Y^E) - I(X^B; Y^{E'})\right]\,.
\end{equation}
Now, from the definition of the capacity of the \ac{MAC} from Alice and Bob to Eve $C_{MAC}(X^A, X^B; Y^E)$ (considering also the \ac{AN} transmitted by Ray) we have 
\begin{equation}
R_{L}<  C_{MAC}(X^A, X^B; Y^E) - \frac{1}{N} I(X^B; Y^{E'})\,.
\end{equation}
Unfortunately, it is hard to to lower bound $I(X^B; Y^{E'})$. We can consider two cases, either Eve is able to decode $\mathcal M^B$ from $Y^{E'}$, or not. Considering the lattice coding, from \cite{Zhu-seminar2014} the first case occurs if $R_{\rm B} \leq C(X^B; Y^{E'})$ where $C(X^B; Y^{E'}) = \max I(X^B; Y^{E'})$ and we have
\begin{equation}
R_{L} \leq C_{MAC}(X^A, X^B; Y^E) - R_{\rm B}\,.
\end{equation}
On the other hand, if Eve is not able to decode $\mathcal M_{\rm B}$ we can only upper bound $R_L$ as follows
\begin{equation}
R_{L} \leq  C_{MAC}(X^A, X^B; Y^E) \,.
\label{leqapp2}
\end{equation}
Since Alice applies random binning, a {\em subset} $(\bar{\mathcal O}_1)^C \subset \bar{\mathcal O}_1^C$ of the {\em non secrecy-outage} event is given by
\begin{equation}
\begin{split}
(\bar{\mathcal O}_1)^C & =  \mathcal S_1 \cup \mathcal S_2 = \\
& \{C_{MAC}(X^A, X^B; Y^E) - R_{\rm B} \leq R_0, \\
& C(X^B; Y^{E'}) \geq R_{\rm B} \}\, \cup  \\
& \{C_{MAC}(X^A, X^B; Y^E) \leq R_0, C(X^B; Y^{E'}) \leq R_{\rm B}\} \,.
\end{split}
\end{equation}
Therefore (\ref{Pout}) becomes
\begin{equation}
P_{\rm out}  \leq 1 - ({\rm P}[\mathcal S_1, \mathcal O_2^C] + {\rm P}[\mathcal S_2, \mathcal O_2^C])\,. 
\end{equation}

\paragraph*{Second Phase} In the second phase we aim at preventing Eve from getting any information on both $\mathcal M_{\rm A}$ and $\mathcal M_{\rm B}$. Alice and Bob transmit \ac{AN} and Ray will employ random binning. Note that we cannot utilize the randomness of the first phase for secrecy purposes, since Eve may have decoded the randomness of the first phase (while not being able of getting any information on the secret message) and can exploit this knowledge in the second phase. This is why we need a second random binning process. Therefore we have a secrecy outage event when 
\begin{equation}
C(X^R; \bar{Y}) \geq R_R\,,
\end{equation}
where $\bar{Y}$ is the signal received by Eve in the second phase. 

\subsection{Rayleigh Fading Scenario}

We now derive the close form expression of the secrecy outage probability bound for the case in which all links with Eve are characterized by Rayleigh fading. In particular, let $1/\lambda_{\rm A}$, $1/\lambda_{\rm B}$, and $1/\lambda_{\rm R}$ be the average \ac{SNR} of the links between Eve and Alice, Bob and Ray, respectively.

Denoting the scalars $q_{\rm A} = |Q_{\rm A}|^2$, $q_{\rm B} = |Q_{\rm B}|^2$, $q_{\rm R} = |Q_{\rm R}|^2$, we have $C_{MAC}(X^A, X^B; Y^E) = \log_2 \left(1 + \frac{q_{\rm A} + q_{\rm B}}{1 +  q_{\rm R}} \right)$ and
\begin{equation}
\begin{split}
\mathcal S_1 = &  \left\{ \log_2 \left(1 + \frac{q_{\rm A} + q_{\rm B}}{1 +  q_{\rm R}} \right)  \leq  R_0 + R_{\rm B}, \right. \\
& \left.  \log_2\left(1 + \frac{q_{\rm B}}{1 + q_{\rm R}}\right) \geq R_{\rm B}  \right\} =  \\
& \{ q_{\rm A} + q_{\rm B}  \leq \mu(1+q_{\rm R}) ,  q_{\rm B}   \geq \nu (1 +q_{\rm R}) \},   \\
 \end{split}
\end{equation}
with $\mu = 2^{R_0+R_{\rm B}}-1$ and $\nu = 2^{R_{\rm B}}-1$, and analogously
\begin{equation}
\begin{split}
 \mathcal S_2 = & \{ q_{\rm A} + q_{\rm B} - \mu' q_{\rm R} \leq \mu',  q_{\rm B}  - \nu q_{\rm R} \leq \nu \}   \\
 = &  \left\{ q_{\rm R} \geq \max \left(\frac{ q_{\rm A} + q_{\rm B} - \mu'}{\mu'}; \frac{ q_{\rm B} -\nu}{\nu}\right) \right\},
\end{split}
\end{equation}
with $\mu' = 2^{R_0}-1$. 

For the second phase we have $C(X^R; \bar{Y}) = \log_2 \left(1 + \frac{q_{\rm R}}{1 + q_{\rm A} + q_{\rm B} }\right)$, and 
\begin{equation}
\begin{split}
 \mathcal O_2^C = & \left\{\log_2 \left(1 + \frac{q_{\rm R}}{1 + q_{\rm A} + q_{\rm B} }\right) \leq R_{\rm R}\right\} = \\
& \{q_{\rm R} - \phi (q_{\rm A} + q_{\rm B}) \leq \phi\} \\
& \left\{q_{\rm A} + q_{\rm B} \geq \frac{q_{\rm R}}{\phi}-1 \right\}, 
\end{split}
\end{equation}
with $\phi = 2^{R_{\rm R}} -1$. Hence we  have
\begin{equation}
{\rm P}[\mathcal O_1^C, \mathcal O_2^C] = {\rm P}[\mathcal S_1, \mathcal O_2^C] + {\rm P}[\mathcal S_2, \mathcal O_2^C]\,. 
\end{equation}

We observe that $\mathcal S_1 \mathcal O^C_2 \neq \emptyset$ if 
\begin{equation}
\frac{q_{\rm R}}{\phi}-1\leq \mu q_{\rm R} + \mu\,,
\end{equation}
which always occurs if $\mu\phi \geq 1$, while it requires
\begin{equation}
q_{\rm R} \leq \frac{\phi + \mu \phi}{1 - \mu\phi }=\gamma  \quad \mbox{if  $\mu\phi < 1$}\,.
\end{equation}
Moreover, condition $\mathcal O_2^C$ has no effect if 
\begin{equation}
\frac{q_{\rm R}}{\phi}-1\leq \nu q_{\rm R} + \nu\,,
\end{equation}
which always occurs if $\nu\phi \geq 1$, and which occurs when
\begin{equation}
q_{\rm R} \leq \frac{\phi + \nu \phi}{1 - \phi \nu} = \delta, \quad \mbox{if  $\nu\phi < 1$}\,.
\end{equation}

\begin{figure*}
\small

\begin{equation}
\begin{split}
Z(A, B, C, D, C', D', E, F)&  = \lambda_{\rm A}\lambda_{\rm B}\lambda_{\rm R}  \int_{q_{\rm R} = A}^{B} \int_{q_{\rm B} = D q_{\rm R}+D'}^{C q_{\rm R} + C'} \int_{q_{\rm A} = E(q_{\rm R}/\phi -1-q_{\rm B})}^{F q_{\rm R}+F-q_{\rm B}} 
 e^{-\lambda_{\rm B} q_{\rm B}} e^{-\lambda_{\rm A} q_{\rm A}} e^{-\lambda_{\rm R} q_{\rm R}} dq_{\rm A} dq_{\rm B} dq_{\rm R} =   \\
&   -\lambda_{\rm B}\lambda_{\rm R}  \int_{q_{\rm R} = A}^{B} e^{-\lambda_{\rm R} q_{\rm R}}\int_{q_{\rm B} = D q_{\rm R}+D'}^{C q_{\rm R} + C'}   e^{-\lambda_{\rm B} q_{\rm B}}    
\left[e^{-\lambda_{\rm A} [Fq_{\rm R} + F - q_{\rm B}]} - e^{-\lambda_{\rm A} E(q_{\rm R}/\phi -1-q_{\rm B})}\right]  dq_{\rm B} dq_{\rm R} 
\end{split}
\label{defZ}
\end{equation}

\begin{equation}
Z_{{\rm x},1}(A, B, C, D,C',D', E, F)=   \frac{\lambda_{\rm B}\lambda_{\rm R}e^{-\lambda_{\rm A} F}}{\lambda_{\rm A} - \lambda_{\rm B}} \left\{   
 \frac{e^{(\lambda_{\rm A} -\lambda_{\rm B}){\rm x}' }}{(\lambda_{\rm A} - \lambda_{\rm B}){\rm x}-\lambda_{\rm R}-\lambda_{\rm A} F}  \left[e^{((\lambda_{\rm A} - \lambda_{\rm B}) {\rm x}-\lambda_{\rm R}-\lambda_{\rm A} F)B} - e^{((\lambda_{\rm A} - \lambda_{\rm B}) {\rm x}-\lambda_{\rm R}-\lambda_{\rm A} F)A} \right] \right\}
\end{equation}
\begin{equation}
X_{{\rm x},1}(A, B, C, D,C',D', E, F)=   \frac{\lambda_{\rm B}\lambda_{\rm R}e^{  \lambda_{\rm A} E}}{\lambda_{\rm A} E - \lambda_{\rm B}}  \left\{   
 \frac{e^{(\lambda_{\rm A}E -\lambda_{\rm B}){\rm x}' }}{(\lambda_{\rm A} E- \lambda_{\rm B}){\rm x}-\lambda_{\rm R}-\frac{\lambda_{\rm A}E}{\phi}}  \left[e^{((\lambda_{\rm A}E - \lambda_{\rm B}) {\rm x}-\lambda_{\rm R}-\frac{\lambda_{\rm A}E}{\phi})B} - e^{((\lambda_{\rm A} E- \lambda_{\rm B}){\rm x}-\lambda_{\rm R}-\frac{\lambda_{\rm A}E}{\phi})A} \right] \right\}
\end{equation}

\begin{equation}
\begin{split}
Z_{2}(A, B, C, D,C',D', E, F)= \frac{\lambda_{\rm B}\lambda_{\rm R}e^{-\lambda_{\rm A} F}(C-D)}{-(\lambda_{\rm R} + \lambda_{\rm A}F)^2} \{e^{-(\lambda_{\rm R} + \lambda_{\rm A}F)B}[(\lambda_{\rm R} + \lambda_{\rm A}F)(B+\frac{C'-D'}{C-D}) +1] \\
 -  e^{-(\lambda_{\rm R} + \lambda_{\rm A}F)A}[(\lambda_{\rm R} + \lambda_{\rm A}F)(A+\frac{C'-D'}{C-D})+1]\}
\end{split}
\end{equation}
\begin{equation}
\begin{split}
X_{2}(A, B, C, D, E, F)= \frac{\lambda_{\rm B}\lambda_{\rm R}e^{\lambda_{\rm A} E}(C-D)}{-(\lambda_{\rm R} + \lambda_{\rm A}E/\phi)^2} \{e^{-(\lambda_{\rm R} + \lambda_{\rm A}E/\phi)B}[(\lambda_{\rm R} + \lambda_{\rm A}E/\phi)(B+\frac{C'-D'}{C-D})+1] -  \\
e^{-(\lambda{\rm R} + \lambda_{\rm A}E/\phi)A}[(\lambda_{\rm R} + \lambda_{\rm A}E/\phi)(A+\frac{C'-D'}{C-D})+1]\}
\end{split}
\end{equation}

\begin{equation}
Z_{{\rm x},2}(A, B, C, D,C',D', E, F)=   \frac{\lambda_{\rm B}\lambda_{\rm R}e^{-\lambda_{\rm A} F}e^{(\lambda_{\rm A} -\lambda_{\rm B}){\rm x} }}{\lambda_{\rm A} - \lambda_{\rm B}} (B-A)\,, \quad X_{{\rm x},2}(A, B, C, D,C',D', E, F)=   \frac{\lambda_{\rm B}\lambda_{\rm R}e^{  \lambda_{\rm A} E}e^{(\lambda_{\rm A}E -\lambda_{\rm B}){\rm x} }}{\lambda_{\rm A} E - \lambda_{\rm B}}  (B-A)\,,  
\end{equation}
\begin{equation}
Z_{\rm x}(A, B, C, D,C',D', E, F)=\begin{cases}
Z_{{\rm x},1}(A, B, C, D,C',D', E, F) & \mbox{ if } (\lambda_{\rm A} - \lambda_{\rm B}){\rm x} -\lambda_{\rm R}-\lambda_{\rm A} F \neq 0 \\
Z_{{\rm x},2}(A, B, C, D,C',D', E, F) & \mbox{ if }(\lambda_{\rm A} - \lambda_{\rm B}){\rm x}-\lambda_{\rm R}-\lambda_{\rm A} F = 0 \\
\end{cases}\,, \quad {\rm x} = {\rm C}, {\rm D}
\end{equation}
\begin{equation}
X_{\rm x}(A, B, C, D,C',D', E, F)=\begin{cases}
X_{{\rm x},1}(A, B, C, D,C',D', E, F) & \mbox{ if }(\lambda_{\rm A} E- \lambda_{\rm B}){\rm x}-\lambda_{\rm R}-\frac{\lambda_{\rm A}E}{\phi} \neq 0 \\
X_{{\rm x},2}(A, B, C, D,C',D', E, F) & \mbox{ if }(\lambda_{\rm A} E- \lambda_{\rm B}){\rm x}-\lambda_{\rm R}-\frac{\lambda_{\rm A}E}{\phi} = 0 \\
\end{cases}\,, \quad {\rm x} = {\rm C}, {\rm D}
\end{equation}
\begin{equation}
T_1(A, B, C, D, E, F)=\begin{cases}
Z_{\rm C}(A, B, C, D,C',D', E, F) - Z_{\rm D}(A, B, C, D,C',D', E, F) & \mbox{ if } \lambda_{\rm A} - \lambda_{\rm B} \neq 0\,, \\
Z_2(A, B, C, D,C',D', E, F) & \mbox{ if } \lambda_{\rm A} - \lambda_{\rm B} = 0  
\end{cases}
\end{equation}
\begin{equation}
T_2(A, B, C, D,C',D', E, F)=\begin{cases}
X_{\rm C}(A, B, C, D,C',D', E, F) - X_{\rm D}(A, B, C, D,C',D', E, F) & \mbox{ if } \lambda_{\rm A}E - \lambda_{\rm B} \neq 0\,, \\
X_2(A, B, C, D,C',D', E, F) & \mbox{ if } \lambda_{\rm A}E - \lambda_{\rm B} = 0  
\end{cases}
\end{equation}
\begin{equation}
Z(A, B, C, D,C',D', E, F) = T_2(A, B, C, D,C',D', E, F) - T_1(A, B, C, D,C',D', E, F)\,.
\label{finalZ}
\end{equation}
\end{figure*}

Considering the definition of $Z(A, B, C, D, C', D', E, F, F)$ in (\ref{defZ})-(\ref{finalZ}) reported in the next page, Therefore we have the following cases
\begin{enumerate}
\item {\bf Case $\mu\phi \geq 1$ and $\phi \nu <1$}. This will lead to ${\rm P}[\mathcal S_1, \mathcal O_2^C] = Z(0, \delta, \mu, \mu, \nu,\nu,0,\mu) + Z(\delta, \infty, 1/\phi, -1,\nu,\nu, 1,\mu) + Z(\delta, \infty, \mu, \mu,1/\phi,-1, 0,\mu)$.
\item {\bf Case $\mu\phi \geq 1$ and $\phi \nu \geq1$}. This will lead to ${\rm P}[\mathcal S_1, \mathcal O_2^C] = Z(0, \infty, \mu,\mu,\nu,\nu,0,\mu)$.
\item {\bf Case $\mu\phi < 1$}. This will imply $\phi \nu <1$ and will lead to ${\rm P}[\mathcal S_1, \mathcal O_2^C] =  Z(0, \delta, \mu,\mu,\nu,\nu,0,\mu) + Z(\delta, \gamma, 1/\phi, -1,\nu,\nu,1,\mu)+ Z(\delta, \gamma, \mu, \mu,1/\phi,-1, 0,\mu)$.
\end{enumerate}

We observe that $\mathcal S_2 \mathcal O^C_2 \neq \emptyset$ if either $\mu'\phi \geq 1$ and $\nu\phi \geq 1$, or $\mu'\phi < 1$, $\nu\phi \geq 1$ and 
\begin{equation}
q_{\rm R} \leq \frac{\phi + \mu' \phi}{1 -  \mu'\phi}=\gamma'\,,
\end{equation}
or $\mu'\phi \geq 1$, $\nu\phi < 1$ and $q_{\rm R} \leq \delta$, or $\mu'\phi < 1$, $\nu\phi < 1$ and $q_{\rm R} \leq \min\{\delta, \gamma'\}$.

Moreover, condition $\mathcal O_2^C$ has no effect if $q_{\rm R} < \phi$. Lastly, if $\mu' \leq \nu$, $\mathcal S_2$ reduces to $\mathcal S_2 = \{q_{\rm A} + q_{\rm B} - \mu'q_{\rm R} \leq \mu\}$ (the other bound has no effect). Therefore we have the following cases
\begin{enumerate}
\item {\bf Case $\mu'\phi \geq 1$,  $\mu' > \nu$ and $\phi\nu \geq 1$}. This will lead to ${\rm P}[\mathcal S_2, \mathcal O_2^C]  = Z(0,\phi,\nu,\nu,0,0,0,\mu') + Z(\phi,\infty,1/\phi, -1, 0, 0,1,\mu') + Z(\phi,\infty,\nu,\nu,1/\phi, -1, 0,\mu')$.
\item {\bf Case $\mu'\phi \geq 1$,  $\mu' > \nu$ and $\phi\nu < 1$}. This will lead to ${\rm P}[\mathcal S_2, \mathcal O_2^C]  = Z(0,\phi,\nu,\nu,0,0,0,\mu') + Z(\phi,\delta,1/\phi, -1, 0, 0,1,\mu') + Z(\phi,\delta,\nu,\nu,1/\phi, -1, 0,\mu')$.
\item {\bf Case $\mu'\phi \geq 1$ and $\mu' \leq \nu$}. This will lead to ${\rm P}[\mathcal S_2, \mathcal O_2^C]  =  Z(0,\phi,\mu',\mu',0,0,0,\mu') + Z(\phi,\infty,1/\phi,-1,0 ,0,1,\mu')+ Z(\phi,\infty,\mu', \mu',1/\phi,-1,0,\mu')$.
\item {\bf Case $\mu'\phi < 1$ and $\mu' > \nu$}. This will lead to ${\rm P}[\mathcal S_2, \mathcal O_2^C]  = Z(0,\phi,\nu,\nu,0,0,0,\mu') + Z(\phi,\min\{\gamma',\delta\},1/\phi,-1,0,0,1,\mu') + Z(\phi,\min\{\gamma',\delta\},\nu, \nu,1/\phi,-1,0,\mu')$.
\item {\bf Case $\mu'\phi < 1$ and $\mu' \leq \nu$}. This will lead to ${\rm P}[\mathcal S_2, \mathcal O_2^C] = Z(0,\phi,\mu',\mu',0,0,0,\mu') + Z(\phi,\gamma',1/\phi,-1,0,0,1,\mu') + Z(\phi,\gamma',\mu',\mu',1/\phi,-1,0,\mu')$.
\end{enumerate}

\begin{figure}
\centering
\includegraphics[height=6.5cm]{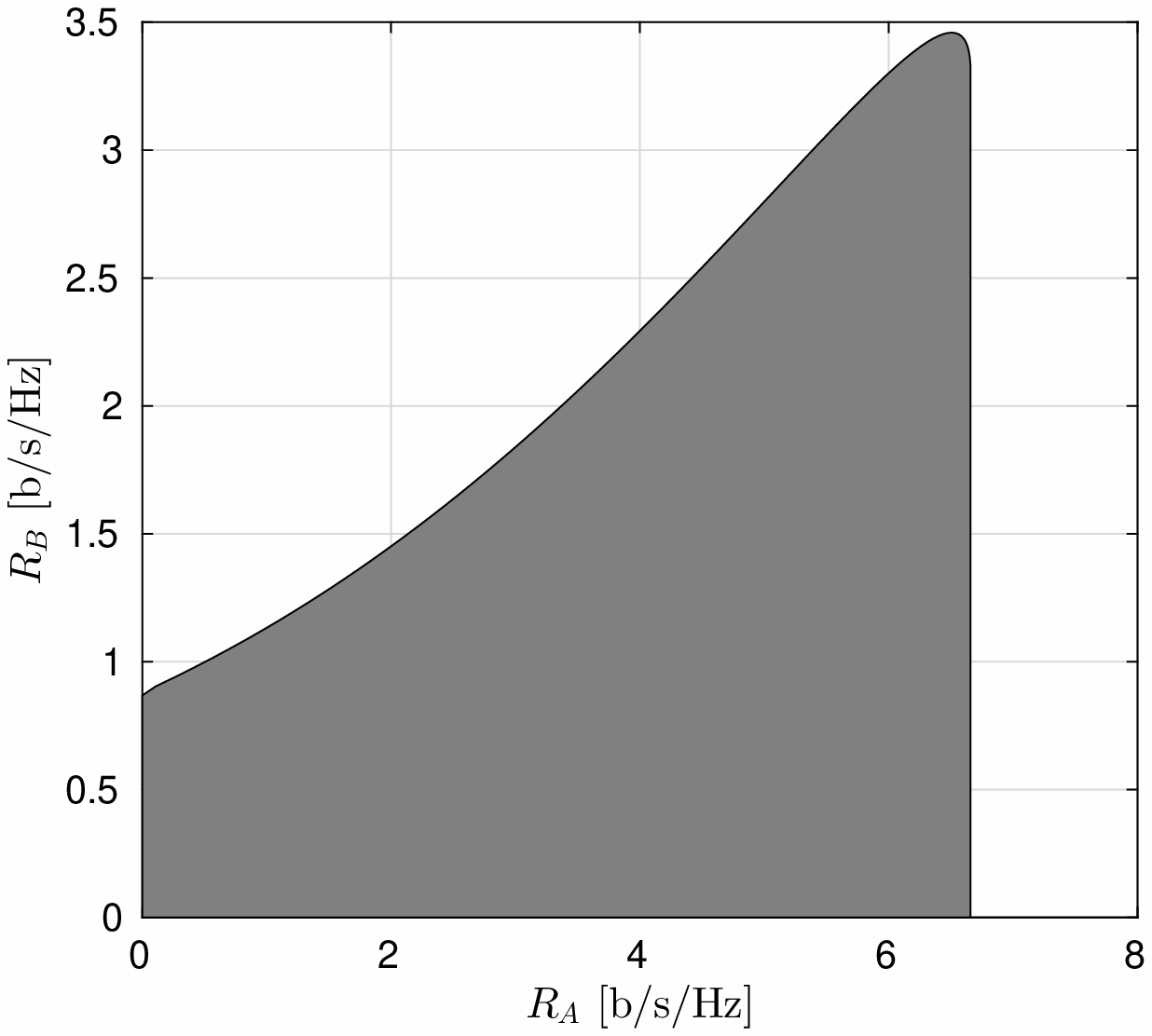}
\caption{Achievable region of $(R_{\rm A}, R_{\rm B})$ couples.}
\label{fig0}
\end{figure}

\section{Numerical Results}

In order to show an example of performance of the considered system, we consider $\delta_{\rm A} = 20$~dB, $\delta_{\rm B} = 10$~dB. We first have spanned the value of $\beta_{\rm A}/\beta_{\rm B}$ and found the achievable region of $(R_{\rm A}, R_{\rm B})$ couples, as shown in Fig. \ref{fig0}.

\begin{figure}
\centering
\includegraphics[height=6.5cm]{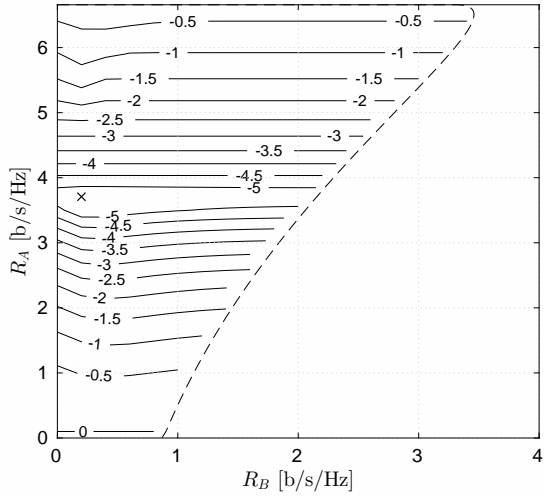}
\caption{Contour plot of $\log_{10} \bar{P}_{\rm out}$ as a function of $R_{\rm A}$ and $R_{\rm B}$. Only values of $R_{\rm A}$ and $R_{\rm B}$ allows decodability conditions at Bob and secrecy at Ray are considered. The couple of $(R_{\rm A}, R_{\rm B})$ providing the minimum value of $\bar{P}_{\rm out}$ is shown with a cross.}
\label{fig1}
\end{figure}

Then we set $\lambda_{\rm A} = \lambda_{\rm R} = 1$, and $\lambda_{\rm B} = 2$, and $R_{\rm S} = 0.1$~bit/s/Hz, and $R_{\rm R} = 7$ bit/s/Hz. Fig. \ref{fig1} shows the bound on the secrecy outage probability (in a log scale) as a function of $R_{\rm A}$ and $R_{\rm B}$. Note that we have only shown the values of $\bar{P}_{\rm out}$ in correspondence of valid couples $R_{\rm A}$ and $R_{\rm B}$ that ensure the decodability conditions for the secret message at Bob, while not leaking any information to Ray. We also identify the couple $(R_{\rm A}, R_{\rm B})$ providing the minimum $\bar{P}_{\rm out}$, which is about $10^{-5}$.

In general, we observe that increasing $R_{\rm A}$ reduces the chances of leaking information to Eve in the first phase (since we can increase $R_0$). On the other hand, since Ray transmits the secret message at rate $R_{\rm A}$, we potentially have an information leakage in the second phase since the rate of the random message $R_{\rm R}-R_{\rm A}$ is decreased. 

\begin{figure}
\centering
\includegraphics[height=6.5cm]{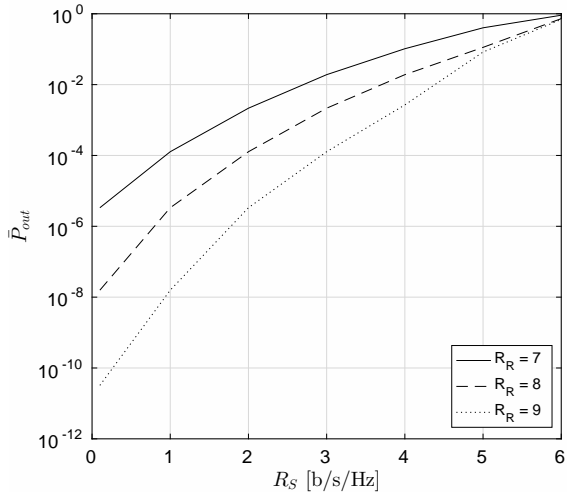}
\caption{Secrecy probability outage bound $\bar{P}_{\rm out}$ vs $R_{\rm S}$ for three values of $R_{\rm R}$. $R_{\rm A}$ and $R_{\rm B}$ have been optimized to minimize $\bar{P}_{\rm out}$.}
\label{fig2}
\end{figure}

Fig. \ref{fig2} shows the secrecy probability outage bound $\bar{P}_{\rm out}$ vs $R_{\rm S}$ for three values of $R_{\rm R}$. $R_{\rm A}$ and $R_{\rm B}$ have been optimized to minimize $\bar{P}_{\rm out}$. We observe that as $R_{\rm S}$ increases, the outage bound increases too, since the random message rates $R_0$ and $(R_{\rm R}-R_{\rm A})$ in the two phases are reduced. Also, increasing $R_{\rm R}$ provides a lower $\bar{P}_{\rm out}$ since it allows to better protect the second phase, increasing the random message rate.

\section{Conclusions}
\balance
We have considered a secure communication scenario where  a secret message must be kept unknown both to a curious but honest device that relays the message to the destination, and to an eavesdropper. While the channel to the relay is known to the legitimate devices, the channel to the eavesdropper is not known, hence a secrecy outage event is possible. Considering \ac{FD} devices and letting Ray transmit \ac{AN}, we have described the protocol that mixes a \ac{SCF} approach and a random binning approach to provide secrecy. Then the secrecy outage probability for a Rayleigh fading scenario has been computed in a close form. Lastly, some numerical results have provided an insight into the main features of the considered system.

\end{document}